# A deep belief network-based method to identify proteomic risk markers for Alzheimer's disease

Ning An
Key Laboratory of Knowledge
Engineering with Big Data of
Ministry of Education,
Hefei University of Technology,
Hefei, China
School of Computer Science and
Information Engineering,
Hefei University of Technology,
Hefei, China
ning.g.an@acm.org

Liuqi Jin
Key Laboratory of Knowledge
Engineering with Big Data of
Ministry of Education,
Hefei University of Technology,
Hefei, China
School of Computer Science and
Information Engineering,
Hefei University of Technology,
Hefei, China
kinuxroot@163.com

Huitong Ding
Key Laboratory of Knowledge
Engineering with Big Data of
Ministry of Education,
Hefei University of Technology,
Hefei, China
School of Computer Science and
Information Engineering,
Hefei University of Technology,
Hefei, China
ding_huitong@163.com

Jiaoyun Yang[†]
Key Laboratory of Knowledge
Engineering with Big Data of
Ministry of Education,
Hefei University of Technology,
Hefei, China
School of Computer Science and
Information Engineering,
Hefei University of Technology,
Hefei, China,
jiaoyun@hfut.edu.cn

Jing Yuan
Department of Neurology,
Peking Union Medical College
Hospital, Chinese Academy of
Medical Sciences,
Beijing, China
yuanjing@pumch.cn

## ABSTRACT

While a large body of research has formally identified apolipoprotein E (APOE) as a major genetic risk marker for Alzheimer's disease, accumulating evidence supports the notion that other risk markers may exist. The traditional Alzheimer's-specific signature analysis methods, however, have not been able to make full use of rich protein expression data, especially the interaction between attributes. This paper develops a novel feature selection method to identify pathogenic factors of Alzheimer's disease using the proteomic and clinical data. This approach has taken the weights of network nodes as the importance order of signaling protein expression values. After generating and evaluating the candidate subset, the method helps to select an optimal subset of proteins that achieved an accuracy greater than 90%, which is superior to traditional machine learning methods for clinical Alzheimer's disease diagnosis. Besides identifying a proteomic risk marker and further reinforce the link between metabolic risk factors and Alzheimer's disease, this paper also suggests that apidonectin-linked pathways are a possible therapeutic drug target.

## 1 Introduction

Alzheimer's disease (AD) is a common form of neurodegenerative disease that is estimated to affect 131 million people worldwide by 2050 [1]. AD accounts for 60% to 70% of cases in all dementia diseases [2, 3], and it can progress for years before symptoms become detectable by conventional means. One approach categorizes AD into two subtypes according to the age at onset: early-onset AD and late-onset AD.

† Corresponding author

Approximately 60% of early-onset AD (EOAD) cases have a history of multiple family members with AD. There are 13% of these cases having autosomal dominant manners. That affects at least three generations [4, 5]. Except for several families with single-gene disorders, most of them present complex situations caused by multiple susceptibility genes and environmental factors. [6-8]. Late-onset AD (LOAD) is a prevalent disease of which the onset age is older than 60 or 65 years. In general, Overall, more than 90% of AD patients are sporadic cases that belong to sporadic LOAD [9]. Studies have shown that there is a genetic factor in LOAD, but no study has identified any causative gene. Indeed, many genetic studies have consistently linked the APOE gene to sporadic LOAD [10-14]. There are other unidentified genetic or environmental factors, as many people with the APOE risk allele (ε4) live into their 90s.

Diagnostic markers should have the ability to reflect pathogenic processes of the AD, including the degeneration of the neurons and synapses [15]. Some studies have identified tau and the 42 amino acid β-amyloid peptide (Aβ42) that met this criterion using traditional methods [16-18]. Variance analysis can compare the differences between AD cases and healthy controls. The Pearson correlation coefficient is computed to assess the correlation between markers and AD. Researchers have calculated sensitivity and specificity by the cutoff value to reflect the proportion of patients with different indicators [19]. These studies, however, require long-term follow-up for clinical neurochemical analyses and assay these two markers weekly. The process is notably time-consuming and complicated. There is a great need for a more effective way to evaluate the diagnostic potential of proteins.

Recently researchers have focused on the development of diagnostic tools for AD [20,55]. The analysis of multiple image modalities is the primary means to understand the pathogenesis and identify the diagnostic markers of AD. For example, researchers have made valuable findings on Magnetic Resonance Imaging (MRI) [21], Positron Emission Tomography (PET) [22], and functional MRI (fMRI) [23]. Several studies have investigated a few Cerebrospinal Fluid (CSF) proteins [24]. Fewer studies have involved serum proteins [25] and using both CSF and serum proteins [26]. Although researchers have worked on improving diagnostic methods over the last several decades, there is still a great need for a fully automated and less subjective method that could detect the disease at earlier stages. This paper deployed a deep belief network (DBN) based method for AD diagnosis using 120 signaling proteins data.

Identifying useful risk markers plays a vital role in determining AD, especially those markers that are detectable in the early stage of the disease. It helps in early intervention to prevent neurodegeneration. Also, during the intervention process, the real-time testing of these markers can help to evaluate the effect of the intervention. Identifying the genes responsible for complex diseases can be very challenging. The most significant limitation of the traditional biological identification methods is the scalability of unique targets. The traditional methods depend on a priori determined hypotheses of different biological significance. There are also statistical methods to identify potential diagnostic biomarkers. Eric Nagele [27] evaluated the differential significance of each biomarker and selected ten autoantibody biomarkers that could effectively differentiate AD, but they did not explain the biological mechanisms that could account for the findings and need to verify the prediction accuracy by constructing other models.

In contrast, the proposed method could select biomarkers that maximize disease diagnosis simultaneously in an automated approach. The proposed method is also able to identify pathogenic factors that are biologically significant, an important breakthrough in understanding the underlying pathology. There is a study that formulizes biomarker identification from signaling proteins expression data as a feature selection problem [28]. In this context, the purpose is to select a group of proteins that is most effective in diagnosing AD. Yang and others consider this problem, and various search strategies could be applied [29]. Besides the capability of dimensional reduction, the improvement of feature selection methods is in need of other aspects, including interpretability, time efficiency, and generalization ability [30].

Researchers have categorized the feature selection methods into filter-based methods, wrapper-based methods, and embedded-based methods, according to how to combine feature search with classification tasks. Filter methods evaluate each feature separately and ignore their interaction. Besides, the feature importance assessment is independent of the classification task. Chaves proposed a filter-based method based on association rule mining algorithm for AD [31].

Computing the correlation between features and labels can help select features in classification tasks. The valuable features are highly correlated with the label while weakly correlated with other features. As a commonly used measure, Spearman's correlation coefficient assesses the degree of correlation between the two features.

Traditional machine learning methods focus on a shallow learning structure that contains only nonlinear transformation. The caveat of the shallow structure is that it cannot represent complex functions, which is limited by sample size and a calculation unit. The complexity of complex classification problems further limits its generalization capacity. There is a need to demonstrate that the non-independence feature representation leads to a better feature set. Neuroscience research has shown that the human visual system uses a multi-layered system for information processing. This clear hierarchy for human perception dramatically reduces the complexity of data processing in the visual system and retains the useful information of the object structure. The success of machine learning largely depends on data representation. Learning representations of the data help extract useful information for classification tasks. These learned representations reveal the nature of the data and can be migratable to other tasks [32]. The practical and theoretical experiences tell us that it is beneficial to use deep architecture to learn complex functions that can represent high-level abstractions. Recently, deep learning-based methods have achieved better performance than other methods for many applications, but little work has utilized it for AD prediction with protein expression data.

Despite numerous studies identifying other AD genetic risk factors, APOE [33, 34] remains the best predictive gene of AD for which there is a scientific consensus. Previous biological studies have hypothesized that adiponectin is also possibly related to AD [35]. This paper proposes a novel feature selection method based on multiple levels of data representation and ranking the feature based on the weights of the deep network. ACRP30 is a protein that is encoded by the ADIPOQ gene in which mutations have been associated with adiponectin deficiency. This paper seeks to determine whether ACRP30 is a potential risk marker for AD diagnosis. Furthermore, we examine whether the proposed deep learning framework can achieve diagnostic accuracy comparable to that of more widely used methods.

## 2    Proposed method

This paper considers the AD diagnosis as a classification problem. The objective is to judge the cognitive status of people according to their signaling protein expression values. This paper proposes a method for disease diagnosis using Deep Belief Network (DBN), which is a probabilistic generative neural network consisting of multiple layers of Restricted Boltzmann Machine (RBM). The preferable performance of DBNs is partly due to the stochastic nature of RBM. This machine could encapsulate a form of robustness to corruption in the representations that DBNs can learn despite the noise during training. A DBN satisfies the good intermediate representation criterion, that is, robustness to the partial destruction of the input [31]. For identifying the proteomic risk marker, this paper also proposed a belief network-based feature selection method using the weights of the network to indicate the importance of the features.

### 2.1    AD diagnosis

As shown in Figure 1, the input data consists of 120 protein expression values that measured from 90 AD cases and 90 non-demented controls (NDC). A DBN model uses this input data as the training dataset. The hidden layer aims to reduce the dimension of the raw data and learn a better representation of the features hidden in the inputs. This model can be useful for any feature extraction task, including biological data, as well as those with classes that are not linearly separable.

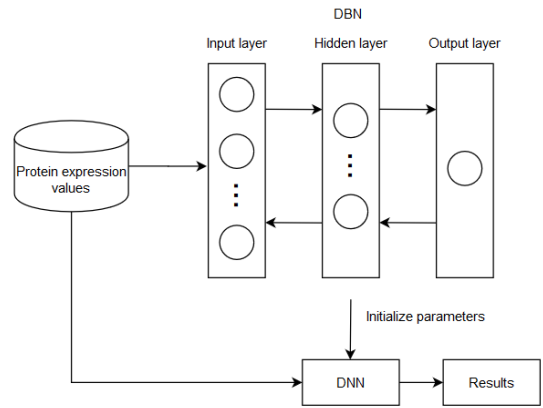

**Figure 1: The proposed framework for AD diagnosis.**

After training the DBN, the proposed method uses the learned weights to initialize a back-propagation neural network (BPNN) for AD diagnosis. BPNN is a multi-layer feedforward network that adopted an error backpropagation training algorithm. It learns and stores many mappings between input-output models without prior mathematic equations that describe this mapping. The learning rule used to adjust the weights and thresholds is the steepest descent method.

This paper interests in two tasks with this model. The first task is to predict whether the participants diagnosed with AD is associated with a particular set of genomic features, such as gene expression data. The method created the training dataset with two classes, where 0.9 and 0.1 indicate AD and NDC respectively, rather than 1 and 0 given the sigmoid activation function.

In the field of machine learning, most methods use learned features to realize their unique targets. These features have no direct relations with the originals. In the era of big biological data, there is an urgent need for effective dimension reduction methods that can form characteristics by selecting the given input attributes. These methods can extract key factors and reduce the necessary number of indexes to solve the problems.

When the training process completes, the weight of the nodes connected each layer on behalf of the contribution to the activation values of upper layer nodes. According to the sum of the absolute value of weights connected input nodes and all nodes in the upper layer, one can choose more important input nodes subsequently. Therefore, this paper adjusts the threshold variably to select a smaller subset of plasma signaling proteins on which the methods could achieve comparative disease diagnostic accuracy.

### 2.2    DBN-based feature selection for identifying proteomic risk marker

This paper proposes a six-step feature selection method based on a designed feature importance criterion and a method for constructing candidate feature subsets. Table 1 describes the details of the algorithm. The weights of the nodes in the input layer and the hidden layer of the trained model serve to indicate the importance of protein expression values in AD diagnosis. After ranking the importance of proteins, important ones are selected in turn to construct candidate feature subsets. The AD prediction method is trained on them to form new models. The performance of these models on different feature subsets is compared by 10-fold cross-validation. The algorithm selects the optimal feature subset that gives the model the highest AD diagnosis performance. At the expense of additional computational burden, the whole procedure can repeat multiple times (in an outer cross-validation loop) to reduce the variance of performance prediction. Extensive experiments show that such an algorithm works well in practice. The training time is tolerable of the compact network size. The experiments suggest that the performance can be improved simply by waiting for bigger datasets to become available.

Table 1. The proposed feature selection method.

| | |
|---|---|
| Input: | Dataset G={X, Y} consists of a sample with protein expression values and label |
| Output: | Feature subset that contains k features |
| Step 1 | **Data preparation** |
| | Divide the dataset G into training dataset {$X_{train}$, $Y_{train}$}, and testing dataset {$X_{train}$, $Y_{train}$}. $X_{train}$ is a matrix in which a column $X_i$ represents the protein expression values of the $i^{th}$ protein in all samples. $Y_{train}$ is the label indicating the cognitive status of samples. |
| Step 2 | **Feature importance criterion** |
| | Choose the weights between input nodes and hidden nodes of the proposed model as the feature importance criterion $R(X_i, Y)$, $1 \leqslant i \leqslant N$. N is the total number of proteins. |
| Step 3 | **Optimal model selection** |
| | Train DBN model on the training dataset, Initialize the NN with the parameters of DBN, Train the NN on the training dataset and adjust the hyperparameters to obtain the best AD diagnosis performance |
| Step 4 | **Candidate feature subset construction** |
| | Sort the features based on the feature importance criterion: $R(X_{r1},Y) \leqslant R(X_{r2},Y) \leqslant ... \leqslant R(X_{rN-1},Y) \leqslant R(X_{rN},Y)$ and construct candidate feature subsets S={$S_1$, $S_2$,..., $S_N$}, in which $S_1$={$X_{r1}$}, $S_2$={$X_{r1}$, $X_{r2}$},..., $S_N$={ $X_{r1}$, $X_{r2}$,..., $X_{rN-1}$, $X_{rN}$}. |
| Step 5 | **Optimal feature subset identification** |
| | For each j, $1 \leqslant j \leqslant N$, Train the NN-based model on feature subset $S_i$ using 10-fold cross-validation. Record the accuracy ACC(j) of the model. End Select the smallest feature set $S_{optimal}$ that has the |
| Step 6 | **Optimal feature subset evaluation** |
| | Train the NN-based model on the training dataset using the feature subset $S_{optimal}$ Test its performance on the testing dataset. |

The causes of AD are still unknown but might have some connection with genes related to cholesterol [56]. This paper fully considers this possibility, and constructs a more compact deep architecture with a smaller number of hidden nodes. This architecture ensures that each hidden unit has a similar impact on the output layer. After training the network, this paper picks out the most crucial attribute of the inputs whose weights to the second hidden layer are the largest. In other words, it is the gene most relevant to AD within the scope of the candidate.

# 3   Results

## 3.1   Dataset

The dataset contains the expression values of 120 blood plasma proteins from 259 plasma samples of AD cases and non-demented controls (NDC). Filter-based arrayed sandwich enzyme-linked immunosorbent assay is used to collect these values [38]. One can download the dataset from http://www.nature.com/nm/journal/v13/n11/suppinfo/nm 1653_S1.html.

## 3.2   Experimental settings

This section presents the experimental settings and results in the AD prediction and feature selection. It carries out 5-fold cross-validation on the dataset to determine the hyperparameters of the proposed model. During the model training, the loss function is calculated according to the result of the forward propagation of each batch of training data. Then the parameters of the model are updated using the gradient descent method. This process uses a learning rate to define the amplitude of each parameter update. If the learning rate is low, it will reduce the speed of network optimization and increase training time. On the contrary, if the learning rate is high, it may cause network parameters to swing back and forth on both sides of the optimal value, resulting in network convergence. The section adopts an effective method to set the learning rate that decays according to the number of iterations. The number of the epoch is set to 100. The model performance on AD phenotype prediction is evaluated by 10-fold cross-validation. As shown in Figure 2, accuracies of three classic machine learning methods, SVM, KNN, and BPNN, are relatively low. These methods are unable to meet the needs of clinical medicine. The proposed method is consistently better than three classic machine learning methods, SVM, KNN, and BPNN.

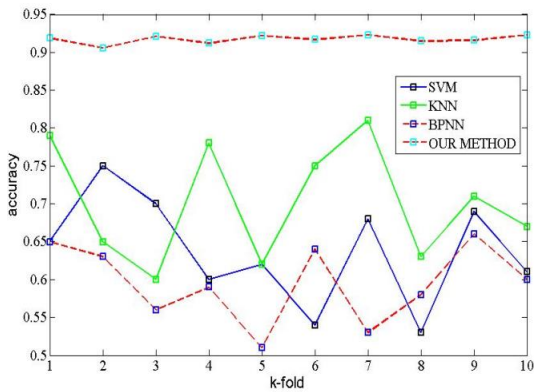

**Figure 2: Performance comparison of the proposed method, SVM, KNN, and BPNN on AD diagnosis.**

*3.2.1 Comparison of the prediction performance on the data before and after using feature selection.* Figure 3 and Figure 4 indicate that the AD diagnosis performance of the proposed model based on the selected 20 signaling proteins is comparable with the performance using 120 signaling proteins. The accuracy of both were greater than 90%, which is significantly better than that of traditional machine learning methods. Therefore, the identification of a small group of signaling proteins in this paper can reduce model complexity and data collection expenses while achieving high diagnostic performance.

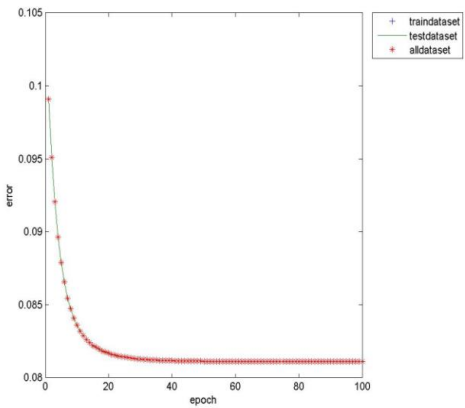

**Figure 3: The AD diagnostic error based on 120 signaling proteins.**

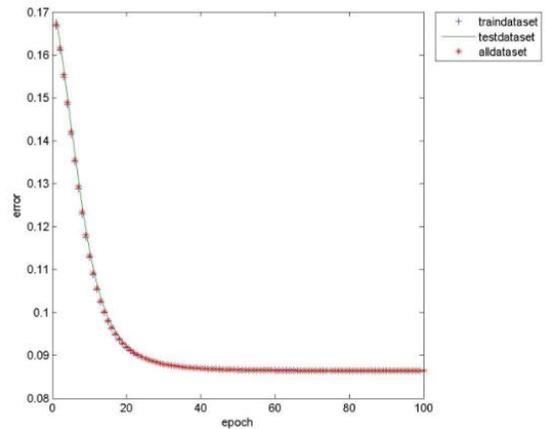

**Figure 4: The AD diagnostic error based on the selected 20 signaling proteins.**

*3.2.2 Identifying ACRP30, a risk marker in AD.* This paper utilizes the hidden layer of the DBN model to learn a high-level feature representation of the signal proteins. The DBN model then uses these features to complete AD diagnosis tasks. On the other hand, this paper can also use these feature representations to reconstruct the raw expression values of signaling proteins. The proposed method uses the reconstruction error to adjust parameters of the prediction model for a better diagnosis performance, which calculated by the difference between the raw expression values and the high-level feature representation. Meanwhile, the optimization of the model for minimum reconstruction error in the training process can help learn a compact and AD-specific feature representation with as much raw information as possible. The optimal parameters of the model can learn the best feature representation with the smallest error. This section uses the average reconstruction error to evaluate the effectiveness of model training and the reliability of feature transformation. If the average reconstruction error of the trained model is small, it indicates that the learned high-level feature representation is credible. On the contrary, if the error is high, it indicates that the trained model cannot learn the high-quality feature representation. The learned model does not fit the data well. This paper minimizes the average reconstruction error by adjusting the number of hidden layer nodes. Figure 5 shows that the average reconstruction error is below 0.05 in the 100 epochs. This low error rate indicates that the model is well trained and can learn desired high-level features. Figure 6 shows AD diagnostic error based on expression values of 120 signaling proteins in the second model.

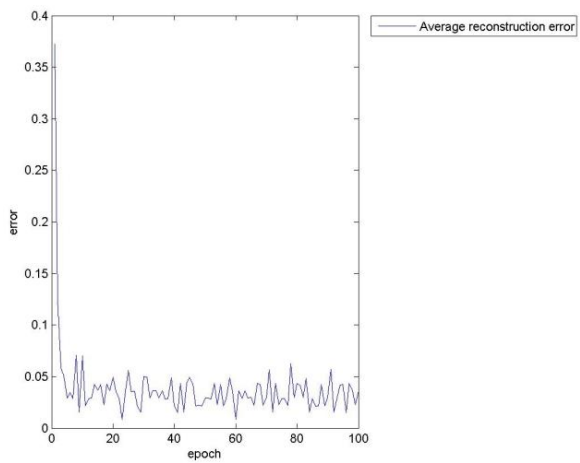

**Figure 5: Average reconstruction error based on 120 signaling proteins in the second model.**

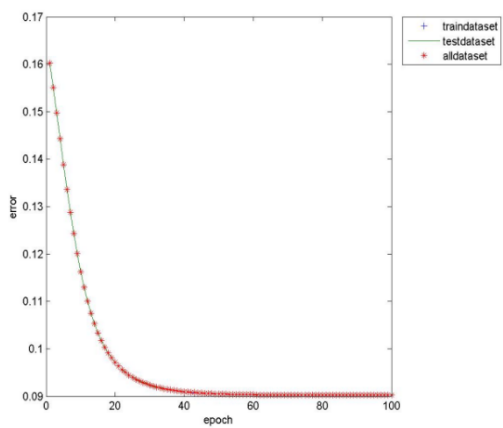

**Figure 6: The AD diagnostic error based on 120 signaling proteins in the second model.**

This paper attempts to identify the relevant biomarkers and determine the etiology of AD by constructing a compact deep model that has little hidden nodes. The model self-learned some features hidden in the original data in an unsupervised way. When the training process of the DBN completes, we gain information on the weights of the connections between the nodes from adjacent layers. These weights indicate the contribution of the nodes to the final prediction results. Figure 7 shows the visualization of the weights between nodes of the input layer and nodes of the hidden layer.

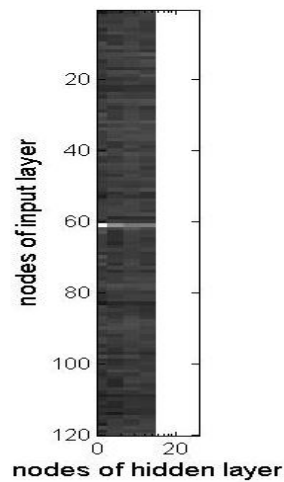

**Figure 7: Visualization of the weights between nodes of the input layer and nodes of the hidden layer.**

For each level, this paper creates a gray-level image by mapping the weights to gray levels such that darker gray levels are for lower weights. A key advantage of having barcode images is that they provide an intuitive, informative, and global view of genomes, from which the importance of various genes becomes immediately apparent.

The horizontal and vertical axes represent the number of first hidden layer nodes and the type of gene expression data, respectively, comprising a 120×5 weight matrix. It is worth noting that the brightest line in the picture is the sixty-first, which corresponding to the ACRP30. This finding indicates that there is a strong association between ACRP30 and AD, which may help establish an early diagnosis marker of AD. These findings are consistent with previous hypotheses, reinforcing the role of adipokines in the pathological mechanism of AD. Figure 8 shows the scores of five proteins, including ACRP30, TIMP-2, HGF, Eotaxin, and IGFBP-1. ACRP30 gets the highest score.

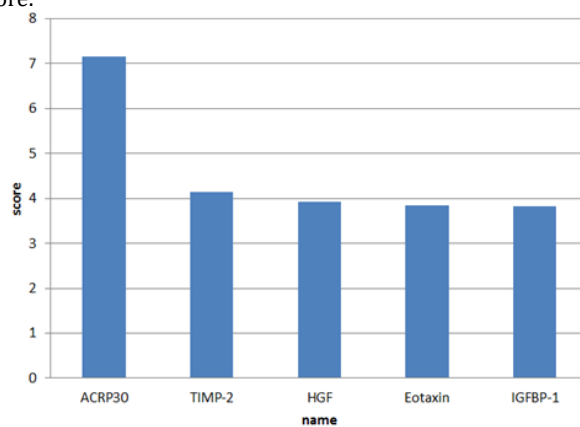

**Figure 8: The calculated importance scores of 5 proteins.**

Correlation is a statistical measure used to assess the degree of linear correlation between two continuous variables. The correlation coefficient, a dimensionless quantity with the values between -1 and 1, is used to indicate the strength of the correlation. The value of the correlation coefficient is 0, indicating that there is no linear correlation between variables. If the correlation coefficient between variables is -1 or 1, it means that there is a perfect linear correlation. Any value between -1 and 1 can represent the strength of the correlation. The stronger the correlation, the closer the absolute value of the correlation coefficient is to 1. This paper calculates the Spearman's correlation coefficient to compare with the proposed feature selection method. Table 2 indicates that the Spearman's correlation coefficient between ACRP30 and AD is 0.048. The two-tailed $P$-value is 0.523. It indicates that the correlation is not significant. Thus, ACRP30 and AD have no linear correlation relationship and suggest that the proposed feature selection method can detect some associations, while the Spearman's correlation coefficients cannot.

**Table 2. Spearman's correlation coefficient between ACRP30 and AD.**

| Correlation | | | ACRP30 | AD |
|---|---|---|---|---|
| Spearman's rho | ACRP30 | Correlation coefficient | 1.000 | .048 |
| | | Sig. (2-tailed) | . | .523 |
| | | N | 180 | 180 |
| | AD | Correlation coefficient | .048 | 1.000 |
| | | Sig. (2-tailed) | .523 | . |
| | | N | 180 | 180 |

## 4 Biological explanation

ACRP30 is the most abundant adipocytokine in plasma in many forms [39]. As a circulating protein, it is synthesized in adipose tissue and there is no clear evidence that it is produced in the brain. ACRP30 plays a central regulatory role in energy homeostasis. Also, some researches have validated ACRP30' neuroprotective role [40, 41]. Researchers suspect that ACRP30 may have multiple roles in neurodegenerative disorders, including AD. Juhyun concludes that it controls the microglia function of the brain [42]. As a surrogate marker, decreased adiponectin/ ACRP30 level could indicate AD pathological changes and links clinical comorbidities, inflammation, and cognitive dysfunction [43]. The process of receptor activation stimulates the intracellular catabolism of ceramide [44]. Many clinical studies have explored its role in AD pathology, but have not reached a consistent conclusion. There are pieces of evidence that AD patients have a lower adiponectin level than healthy people [57]. However, it does not have the prediction ability of dementia progression [45].

In addition to the direct regulation of the disease, ACRP30 also controls the disease indirectly by regulating other factors that affect the pathogenesis. Credible researches have shown that the accumulation of amyloid-β peptide (Aβ) is a critical marker in the pathogenic process of AD. ACRP30 could reduce Aβ generation and accumulation [46]. ACRP30 could also improve insulin resistance in the central nervous system [47], prevent nerve cell apoptosis [48, 49], vascular atherosclerosis, and regulate glycolipid metabolism [50, 51]. In the Japanese population, a scientific paper shows that serum adiponectin level positively correlates with HDL-cholesterol [52]. The clinical study has shown that the high plasma cholesterol level is a risk factor for AD [53]. However, the cholesterol-AD hypothesis presents some difficulties. There is no consensus that the cholesterol level in plasma is indicative of cholesterol metabolism in AD brains. Some studies have reported that there is no relation between serum cholesterol levels and AD. Our results provide counterevidence in support of a link.

Several epidemiological studies have linked obesity to AD [54], but the role of adiposity across the course of cognitive decline is not well-understood to date. Coronary artery disease patients have low plasma adiponectin levels. Moreover, some people have suggested an association between hypoadiponectinemia and carotid atherosclerosis [58]. The present study demonstrates that the plasma adiponectin level is positively related to sex, HDL-cholesterol, and BMI. Therefore, based on the results of this paper, there is a significant indication that cholesterol is an AD risk marker.

ACRP30 has a direct or indirect regulatory effect on the AD pathological process. This paper designs a new feature selection method to support this finding from the perspective of data science. Therefore, it can serve as a useful therapeutic target to alleviate AD manifestations.

## 5 Conclusions

This paper uses a novel method based upon DBN to advance the understanding of AD diagnosis as it relates to blood plasma protein levels. Even more important than the precision of this method is its generalizability. Due to the influence of the structural characteristics of DBN, it is convenient to learn the nature of expression data automatically and promising for medical applications in the era of big data. This paper proposes a feature selection algorithm that ranks the features according to the weight in a deep network. The size of the feature subset can be set variably according to the balance of the diagnostic accuracy and complexity of data collection. Thereby, the potential correlation between ACRP30 and AD has been demonstrated using computational science for the first time and suggests its potential biological significance. The experiments suggest that the proposed method is significantly better than classical machine learning classification methods, including KNN, SVM, and BPNN. One can obtain similar forecasting performance using a subset of features, which significantly reduces the number of indicators that must be collected to predict the disease.

The proposed feature selection method offers the potential to overcome the problems of traditional approaches with feature dimensionality and limited size data sets. This method also simplifies the measurement index required for diagnosis. One could select a subset of factors based on the expected diagnostic accuracy. This paper shows ACRP30 to be a causative factor in AD both by unsupervised and deep learning methods. However, the study also demonstrated that obesity and cholesterol are risk factors for AD. These results enhance the genetic knowledge of AD and point out that apidonectin-linked pathways could be a therapeutic drug target.

## ACKNOWLEDGMENTS

This work was supported in part by the National Key R&D Program of China under Grant No. 2018YFB1003204, CAMS Initiative for Innovative Medicine (CAMS-I2M, No. 2016-I2M-1-004), the Anhui Provincial Key Project of Research and Development Plan under Grant No. 1704e1002221, and the Programme of Introducing Talents of Discipline to Universities ("111 Program") under Grant No. B14025.